\documentstyle[aps,epsfig,amsmath,twocolumn]{revtex}%

\begin{document}
\draft
\tighten
\title{Polarization quantum properties in type-II Optical Parametric
Oscillator below threshold}
\author{Roberta Zambrini$^1$, Alessandra Gatti$^2$, Luigi Lugiato$^2$,
Maxi San Miguel$^1$}
\address{$^1$ Instituto Mediterr\'aneo de Estudios Avanzados,
IMEDEA (CSIC-UIB),\\ Campus Universitat Illes Balears, E-07122
Palma de Mallorca, Spain. }
\address{$^2$INFM, Dipartimento di Scienze CC.FF.MM.,
Universit\'a dell'Insubria, Via Valleggio 11, 22100 Como, Italy}
\maketitle

\centerline{\today}

\begin{abstract}

We study the far field spatial distribution of the  quantum
fluctuations in the transverse profile of the output light beam
generated by a type II Optical Parametric Oscillator below
threshold, including the effects of transverse walk-off.  We study
how quadrature field correlations depend on the polarization. We
find spatial EPR entanglement in quadrature-polarization
components:  For the far field points not affected by walk-off
there is almost complete noise suppression in the proper
quadratures difference of any orthogonal polarization components.
We show the entanglement of the state of symmetric intense, or
macroscopic, spatial light modes. 
We also investigate nonclassical polarization properties in terms
of the Stokes operators. 
We find perfect correlations in
all Stokes parameters measured in opposite far field points in the
direction orthogonal to the walk-off, while locally the field is
unpolarized and we find no polarization squeezing.

\end{abstract}


\section{Introduction}
\label{sect:intro}

Polarization \cite{Schnabel} and transverse spatial
\cite{specialissue} degrees of freedom of light beams interacting
with nonlinear media have been extensively studied in the last
decade.  The selection of special spatial modes of the transverse
profile of a light beam down-converted by a quadratic crystal
provides an interesting example of a  polarization entangled state
\cite{kwiat}. In this kind of experiments \cite{kwiat} the
fluorescence, or rate of photon pairs production, is low (single
photon regime). Recently there has been an increasing interest for
polarization entanglement in continuous variable regimes, where
intense light beams, with high fluxes of photons, are generated.
In this case the detection no longer resolves single photon events
\cite{Korolkova.2002,bowen}. The interest in such macroscopic or
multiphoton systems is partly due to possible applications of
continuous variables in quantum communications \cite{leuch},
quantum information \cite{lloyd}, mapping from light to atomic
media \cite{polzik.99}, and quantum teleportation
\cite{polzik.2000}. Most works on continuous variable regimes
\cite{Korolkova.2002,bowen,leuch,lloyd,polzik.99,polzik.2000} are
concerned only with temporal features of light beams, while our
aim in this paper is to study polarization entanglement between
spatial modes of intense light beams, when intensities are
continuous variables.

Interesting polarization effects arise in type II phase matching,
when a pump field is down-converted in a quadratic crystal in two
orthogonally polarized fields \cite{kwiat}. Parametric
down-conversion (PDC) can be increased using an intense pump pulse
or by means of a resonant optical cavity. The case of an intense
pump is considered in Ref. \cite{OPA}.  In this paper we study the
situation of an optical cavity, that is a Optical Parametric
Oscillator (OPO). In the OPO the fields resonate in a cavity and
therefore an intense laser-like beam is down-converted above
threshold. The cavity enhances the rate of production of photons
pairs, increasing the gain and providing active filtering of the
frequency bandwidth \cite{ou-lu,Oberparleiter}. The spatial
distribution of the quantum fluctuations close to threshold is
dominated by weakly damped modes that become unstable at the
threshold of the OPO, as discussed in Sect.\ref{sect:1}.  This
fluctuating spatial structure, known as `quantum image',  has been
extensively studied for type I OPO, where polarization does not
play any important role \cite{OPOale,Szwaj,roby2}. Here we analyze
the spatial quantum fluctuations for type II phase matching,
considering the polarization and the transverse spatial degrees of
freedom. We also study the effects of the  transverse walk-off
between the orthogonally polarized signal and idler down-converted
fields.

The characterization of the spatial and polarization properties of
the down-converted light is given in two different ways, discussed
in Sect. \ref{sect:2} and in Sect. \ref{sect:3}, respectively. In
Sect.\ref{sect:2} we study Einstein-Podolsky-Rosen (EPR)
correlations \cite{epr} between polarization-quadratures
components. A precedent of macroscopic EPR experiments in OPO are
those of Ref.\cite{ou.92,Zhang}, but they do not refer to spatial
EPR, distinguishing signal and idler by their polarization.
Spatial EPR was theoretically considered in Refs.
\cite{gatti.EPR1,gatti.EPR2} in type I-OPO. We build on these
results by considering the polarization degree of freedom and
including the walk-off effects in our treatment. Our main finding
is that for the intersection points of the far field rings there
is noise suppression much below the standard quantum limit in the
proper quadratures combinations of any orthogonal polarization
components of the critical modes. We show that the entanglement
observed in single photon regime \cite{kwiat} survives for the
multipohoton state generated by an OPO near below threshold.

In Sect.\ref{sect:3} we analyze the issue of macroscopic
polarization entanglement in terms of Stokes operators. These
operators are related to the intensity of different polarization
components of light. There is a recent experimental observation of
polarization squeezing in  three of the four  Stokes parameters
\cite{Bachor.2002}, while macroscopic polarization entanglement in
terms of Stokes parameters has been considered in Ref.
\cite{Korolkova.2002} and demonstrated in Ref. \cite{bowen}.  The
situations considered in
Refs.\cite{Bachor.2002,Korolkova.2002,bowen} do not involve
transverse spatial degrees of freedom.  Quantum properties of
Stokes operators taking into account the spatial transverse
dependence of light seem to be first considered here and in Ref.
\cite{OPA}. We study both the local properties of Stokes
parameters in the transverse far field profile of the
down-converted beam (Sect.\ref{sect:3.local}) and the possibility
of entanglement between spatial far field modes
(Sect.\ref{sect:3.corr}). As a main result we show that there
exists macroscopic polarization entanglement between  the beams
measured at the  intersection points of the far field rings. We
find perfect quantum correlation at zero-frequency in all Stokes
operators.

\section{Input/output relations and far field characteristics}
\label{sect:1}

We consider a type II OPO below threshold, in the undepleted pump
approximation. In this approximation the pump field is described by a fixed classical
variable $A_0$. In the mean field approximation the signal ($\hat A_1$,
ordinary $x$  polarized) and the idler ($\hat A_2$,
extraordinary $y$  polarized) fields obey the following
Heisenberg operator equations \cite{Szwaj}:
\begin{eqnarray}
\nonumber
\partial_t \hat A_1&=& -\gamma_1(1+i\Delta_1-ia_1\nabla^2 +
\rho_1{\partial_y})\hat A_1 +\sqrt{\gamma_1\gamma_2}{A}_0
\hat A_2^\dagger\\
&+&\sqrt{2\gamma_1}\hat A_1^{in}
\label{Eq.A1}\\
\nonumber
\partial_t \hat A_2&=& -\gamma_2(1+i\Delta_2-ia_2\nabla^2 +
\rho_2{\partial_y})\hat A_2 +\sqrt{\gamma_1\gamma_2}{A}_0
\hat A_1^\dagger\\
&+&\sqrt{2\gamma_2}\hat A_2^{in}\label{Eq.A2}
\end{eqnarray}
where $\gamma_i$ are  the cavity linewidths for the signal and idler
fields, $\Delta_i$ are the cavity detunings,  $\nabla^2$ is the two
dimensional transverse Laplacian that models the effect of diffraction in
the paraxial approximation, and $a_i$ are the diffraction strengths.
In type II phase matching a transverse walk-off
arises between the signal and idler fields. It is described by the drift
terms $\rho_i{\partial_y}\hat A_i$.
 $\hat A_i^{in}$ are the the input field operators, describing the
fluctuations of the signal and idler modes entering through the
partially transmitting cavity mirror.

From the Fourier transform:
\begin{eqnarray}\label{fourier}
\hat A(\vec x,t) &=& \int \frac{d^2 \vec k}{2\pi} \int \frac{d\omega}{\sqrt{2\pi}}
e^{i(\vec k\cdot\vec x-\omega t)}\hat A(\vec k,\omega) \\
\hat A^\dagger(\vec x,t) &=& \int \frac{d^2 \vec k}{2\pi} \int \frac{d\omega}{\sqrt{2\pi}}
e^{i(\vec k\cdot\vec x-\omega t)}\hat A^\dagger(-\vec k,-\omega)
\end{eqnarray}
we obtain the following algebraic linear relation,
giving the
intracavity mode operators $\boldsymbol{\hat{\nu}(\vec k,\omega)}$ in
terms of the input fields $\boldsymbol{\hat{\nu}^{in}(\vec k,\omega)}$
\begin{eqnarray}
\boldsymbol{L}\boldsymbol{\hat{\nu}}=\boldsymbol{\Gamma}
\boldsymbol{\hat{\nu}^{in}},
\end{eqnarray}
where we have introduced the operator vectors
\begin{eqnarray}
\boldsymbol{\hat{\nu}}=\left( \begin{array}{c}
\hat A_1(\vec k,\omega)\\ \hat A^{ \dagger}_2(-\vec k,-\omega)
\end{array} \right) , \,
\boldsymbol{\hat{\nu}^{in}}=\left( \begin{array}{c}
\hat A^{in}_1(\vec k,\omega)\\ \hat A^{in \dagger}_2(-\vec k,-\omega)
\end{array} \right) , \,
\end{eqnarray}
and the matrices
\begin{eqnarray}
\boldsymbol{L}&=& \left( \begin{array}{cc}
\gamma_1(1+i\Delta_1(\vec k,\omega))&-\sqrt{\gamma_1\gamma_2}A_0\\
-\sqrt{\gamma_1\gamma_2}A^*_0&\gamma_2(1-i\Delta_2(-\vec k,-\omega))
\end{array} \right)  , \\
\boldsymbol{\Gamma}&=&\left( \begin{array}{cc}
\sqrt{2\gamma_1}&0\\
0&\sqrt{2\gamma_2}
\end{array} \right) ,\,
\end{eqnarray}
with
\begin{eqnarray}\label{eq:detunings}
\Delta_j(\vec k,\omega)=\Delta_j+a_j|\vec k|^2+\rho_j k_y-\omega/\gamma_j.
\end{eqnarray}
Using the input/output relations $\boldsymbol{\hat{\nu}^{out}}=
\boldsymbol{\Gamma}\boldsymbol{\hat{\nu}}-\boldsymbol{\hat{\nu}^{in}}$
\cite{collett84} we obtain the  output fields:
\begin{eqnarray}\label{modes_eq}
\hat A^{out}_{1,2}(\vec k,\omega)=
U_{1,2}(\vec k,\omega)\hat A^{in}_{1,2}(\vec k,\omega)+
V_{1,2}(\vec k,\omega)\hat A^{in}_{2,1}(-\vec k,-\omega).
\end{eqnarray}
The coefficients of the input/output transformation are:
\begin{eqnarray} \label{Eq:U1}
U_1(\vec k,\omega)&=&\frac{2[1-i\Delta_2(-\vec k,-\omega)]}
{[1+i\Delta_1(\vec k,\omega)][1-i\Delta_2(-\vec k,-\omega)]-|A_0|^2}-1\\
\label{Eq:V1}
V_1(\vec k,\omega)&=&\frac{2A_0}
{[1+i\Delta_1(\vec k,\omega)][1-i\Delta_2(-\vec k,-\omega)]-|A_0|^2}
\end{eqnarray}
and $U_2$, $V_2$ are obtained interchanging the indexes $1$ and
$2$ in Eqs. (\ref{Eq:U1}) and Eq. (\ref{Eq:V1}).

Assuming that the input signal and idler fields are in the vacuum
state, it is immediate to obtain the non-vanishing second order
moments of the output fields
\begin{eqnarray}
\label{A1+A1}
\langle\hat A^{out~\dagger}_1(\vec k,\omega)\hat A^{out}_1(\vec k',\omega')\rangle
&=&|V_1(\vec k,\omega)|^2\delta(\vec k-\vec k')\delta(\omega-\omega'),\\
\label{A2+A2}
\langle\hat A^{out~\dagger}_2(\vec k,\omega)\hat A^{out}_2(\vec k',\omega')\rangle
&=&|V_2(\vec k,\omega)|^2\delta(\vec k-\vec k')\delta(\omega-\omega'),\\
\label{A1A2}
\langle\hat A^{out}_1(\vec k,\omega)\hat A^{out}_2(\vec k',\omega')\rangle
&=&U_1(\vec k,\omega)V_2(-\vec k,-\omega)\\\nonumber
\delta(\vec k+\vec k')\delta(\omega+\omega')
\end{eqnarray}
and the corresponding hermitian conjugate ones. From these
moments, quadrature and intensity correlations  can be
analytically calculated for a transversally homogeneous pump
$A_0$. These calculations can be simplified considering the
unitary conditions (see, for instance, Ref. \cite{Szwaj}) and the
symmetries of the system. From Eqs.(\ref{A1+A1})-(\ref{A2+A2}) we
obtain the local Far  Field (FF) intensity. From Eq.(\ref{A1A2})
it is evident that correlations between signal and idler are non
vanishing only between symmetric points of the far field. Such
two-mode correlations are due to  the quadratic form of the
Hamiltonian describing, within a linear approximation, the OPO
below threshold. In the following, and in order to avoid
unphysical divergences \cite{singularity},
 we consider that the fields are integrated over a
detection region $R_{\vec k}$ of area $\sigma$:
\begin{eqnarray} \label{Eq:discr_area}
\hat A^{out}(\vec k)\rightarrow \int_{R_{\vec k}} dk'\hat A^{out}(\vec k').
\end{eqnarray}

At the threshold  for signal generation ($A_0^{thr}=1$), and for a
negative total detuning ($\gamma_1\Delta_1+\gamma_2\Delta_2<0$),
an instability at finite wavenumber and with nonzero frequency
appears, as extensively discussed in Refs.
\cite{gonzalo,ward2001}. The modes that become unstable at
threshold are determined by
 the relations $\Delta_1(\vec k,\omega_H(\vec k))=0$ and
$\Delta_2(\vec k,\omega_H(\vec k))=0$, where
\begin{eqnarray} \label{Eq:wH}
\omega_H(\vec k)&=&\frac{\gamma_1\gamma_2}{\gamma_1+\gamma_2}\cdot
\\\nonumber
&&[\Delta_1-\Delta_2+(a_1-a_2)|\vec k|^2+(\rho_1+ \rho_2)k_y]
\end{eqnarray}
is the frequency that becomes undamped at threshold (Hopf bifurcation).

\begin{figure}[t]
\includegraphics[width=0.4\textwidth,clip=true]{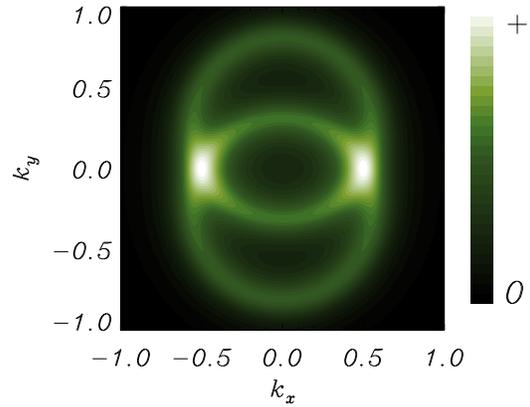}
\caption[Far field OPO intensity]{Far field intensity Eq. (\ref{Eq:It}) for
 $A_0=0.99$. Space is scaled with diffraction strenght $a_1=a_2$ and
time is scaled with cavity decay $\gamma_1=\gamma_2$ \cite{lane}, as
reported in Ref.\cite{OPOale,roby2}.
Other parameters are  $\Delta_1=\Delta_2=-0.25$, $\rho_1=0$, $\rho_2=1$.
The same parameters will be used for all Figures, unless another choice
is specified.
The lower (upper) ring corresponds to the intensity of the field
$1$ ($2$).} \label{ch:5.fig:0}
\end{figure}

The unstable critical modes  lie on
two  rings of the far field given by:
\begin{eqnarray} \nonumber
\gamma_1\Delta_1+\gamma_2\Delta_2+(\gamma_1a_1+\gamma_2a_2)
|\vec k|^2\pm(\gamma_1\rho_1-\gamma_2\rho_2)k_y\\\label{Eq:rings}=0.
\end{eqnarray}
If the relative walk-off $\gamma_1\rho_1-\gamma_2\rho_2$ vanishes,
Eq.(\ref{Eq:rings}) describes a single far field ring. Therefore
in absence of walk-off the signal and the idler far field
distributions are superimposed, and an intense ring is observed.
The two rings of Eq.(\ref{Eq:rings}) are clearly identified in
Fig.\ref{ch:5.fig:0}, where we represent the stationary mean
intensity close to threshold:
\begin{eqnarray} \label{Eq:It}
\langle\hat A^{out~\dagger}_1(\vec k,t)\hat A^{out}_1(\vec k,t)
+      \hat A^{out~\dagger}_2(\vec k,t)\hat A^{out}_2(\vec k,t)\rangle=
\\\nonumber
\frac{\sigma}{2\pi}\int d\omega (|V_1(\vec k,\omega)|^2+
|V_2(\vec k,\omega)|^2).
\end{eqnarray}

This figure is similar to the well-known experimental image
obtained  when there is no cavity, in spontaneous parametric down
conversion \cite{kwiat,OPA}. We want to point out that the cavity
introduces fundamental differences: in particular in the cavity
case there is a threshold above which a pattern appears in the
transverse profile of the fields. The modulus of the wave-vector
of such pattern at threshold is identified in the noisy
$precursor$ below threshold \cite{JeffriesWeisenfeld}. The
existence of a selected wave-number is an effect of the optical
cavity, its value being determined by the cavity detuning
\cite{gonzalo}.

The main contribution to the integrals in Eq. (\ref{Eq:It}) are
the most intense
frequencies components given by Eq. (\ref{Eq:wH}).
Hence the intensity at the Hopf frequency
\begin{eqnarray}\nonumber
\langle\hat A^{out~\dagger}_1(\vec k,\omega_H(\vec k))
\hat A^{out}_1(\vec k,\omega_H(\vec k))+\\\nonumber
\hat A^{out~\dagger}_2(\vec k,\omega_H(\vec k))
\hat A^{out}_2(\vec k,\omega_H(\vec k))\rangle
\end{eqnarray}
is very similar to the one of Fig. \ref{ch:5.fig:0}. This picture
of the FF shows that the intensity reaches the highest value at
the intersection points of the two rings, where ordinary and
extraordinary fields are superimposed. From Eq.(\ref{Eq:rings}) it
is immediate to obtain the coordinates of these crossing points:
\begin{eqnarray} \label{Eq:kx.crit}
\pm \vec k_H=(\pm k_x^c,0) \textrm{, with }k_x^c=\sqrt{ \frac{-\gamma_1\Delta_1-\gamma_2\Delta_2}
{\gamma_1a_1+\gamma_2a_2}}.
\end{eqnarray}
In our calculations we will also consider the FF modes on the rings for which
 the influence  of the walk-off is stronger. These are the four
points of intersection of the two rings and the  $y$ axis, with
ordinates:
\begin{eqnarray}  \label{Eq:ky.crit}
&&\pm k_{y\pm}^c=\frac{1}
{2(\gamma_1a_1+\gamma_2a_2)}\left[\pm(\gamma_1\rho_1-\gamma_2\rho_2)
\right.\\\nonumber
&&\left.\pm\sqrt{
(\gamma_1\rho_1-\gamma_2\rho_2)^2-4(\gamma_1\Delta_1+\gamma_2\Delta_2)
(\gamma_1a_1+\gamma_2a_2)}\right].
\end{eqnarray}
We define the two external points by
$\pm \vec k_{V}=(0,\pm k_{y+}^c)$,
with $k_{y+}^c$ obtained from Eq. (\ref{Eq:ky.crit}) with both $+$
signs.

For simplicity in the following  we shall omit the label $out$,
indicating with $\hat A_{1,2}$  the output fields, described by
Eqs. (\ref{A1+A1})-(\ref{A1A2}).

\section{Spatial EPR entanglement between quadrature-polarization field
components}
\label{sect:2}
The vectorial field is a superposition of linearly polarized
components:
\begin{eqnarray}\label{tot-field}
\hat {\vec A}=\hat A_1\vec e_x+\hat A_2 \vec e_y.
\end{eqnarray}
By means of a wave retarder
\begin{eqnarray}\label{eq:W}
W=\left( \begin{array}{cc}
1&0\\
0&e^{i\Gamma}
\end{array} \right) \,
\end{eqnarray}
and a polarization rotator
\begin{eqnarray}\label{eq:P}
P=\left( \begin{array}{cc}
~\cos\Theta&\sin\Theta \\
-\sin\Theta&\cos\Theta
\end{array} \right) \, ,
\end{eqnarray}
we can obtain a field ${\hat{\vec A}_{\Gamma\Theta}}=WP\hat{\vec A}$
in any polarization state
\begin{eqnarray}\nonumber
\hat {\vec A}_{\Gamma\Theta}=
( \hat A_1\cos\Theta+\hat A_2 e^{i\Gamma}\sin\Theta ) \vec e_x+\\\nonumber
(-\hat A_1\sin\Theta+\hat A_2 e^{i\Gamma}\cos\Theta ) \vec e_y.
\end{eqnarray}

With  a linear polarizer
\begin{eqnarray}
L=\left( \begin{array}{cc}
1&0\\
0&0
\end{array} \right) \, ,
\end{eqnarray}
we can select a field polarization component, and,
integrating over a detection region $R_{\vec k}$,
we  obtain
\begin{eqnarray}\nonumber
\hat A_{\Gamma\Theta}(\vec k,t)=\int_{R_{\vec k} }
d\vec k'(\hat A_1(\vec k',t) \cos\Theta+\hat A_2(\vec k',t)
e^{i\Gamma}\sin\Theta ).
\end{eqnarray}

By  homodyne detection we can select a quadrature component
of this polarization component.
We define the quadrature $\Psi$ by
\begin{eqnarray}\label{quadr.eq}
A_{\Gamma\Theta}^\Psi(t)=A_{\Gamma\Theta}(t)e^{i\Psi}+
(A_{\Gamma\Theta})^\dagger(t) e^{-i\Psi}.
\end{eqnarray}

In any FF point, the arbitrary quadrature-polarization component
(\ref{quadr.eq}) has vanishing mean value and a spectral variance
which depends only on $\Theta$, but it is independent of the choice of the phase
factors $\Psi$ and $\Gamma$. Integrating over a detection region
$R_{\pm\vec k}$ of area $\sigma$, much smaller of the variation scale
 of $U_i$ and $V_i$,  we obtain
\begin{eqnarray}\label{loc_squeez}
&&\int dt e^{i\omega t}\langle\hat A_{\Gamma\Theta}^\Psi(\vec k,t)
A_{\Gamma\Theta}^\Psi(\vec k,0)\rangle=\\ \nonumber
&&\sigma
[1+\cos^2\Theta(|V_1(\vec k,\omega|^2+|V_1(\vec k,-\omega|^2)+
\\ \nonumber
&&\sin^2\Theta(|V_2(\vec k,\omega|^2+|V_2(\vec k,-\omega)|^2],
\end{eqnarray}
where $\sigma$ fixes the shot noise level. Therefore, in any far field point
the variance of an arbitrary quadrature-polarization component is above the
shot noise level $\sigma$. We observe that if we place the detectors
in the FF line  $k_y=0$ the variance results independent of all angles,
including $\Theta$. We find
\begin{eqnarray}\nonumber
&&\int dt e^{i\omega t}\langle\hat A_{\Gamma\Theta}^\Psi((k_x,0),t)
A_{\Gamma\Theta}^\Psi((k_x,0),0)=\\ \nonumber
&&\sigma
[1+|V_1((k_x,0) ,\omega|^2+|V_1((k_x,0) ,-\omega)|^2],
\end{eqnarray}
In other words, for $k_y=0$, the level of fluctuations
is independent of the choice of the polarization state
and quadrature, depending only on the position $k_x$.

Next, we  consider the correlations between the field components detected
from two spatially separated FF points. Such
correlations are not vanishing only for \emph{symmetric} FF points $\vec k$ and
$-\vec k$. We will show that the correlations between quadratures
measured from these two positions of the transverse plane show EPR entanglement.
A two mode state is here defined to be  EPR entangled if for
 two orthogonal quadratures $\hat X_i$ and $\hat Y_i$ in each mode $i$
($i=1,2$) the conditional variances $V^-_{cond}(\hat X_1|\hat X_2)$
and $V^+_{cond}(\hat Y_1|\hat Y_2)$ are \emph{both} less than
$1$, as discussed in Ref.\cite{leuch}. The conditional variance is defined
by
\begin{eqnarray}\label{EPR.ent.leuch}
V^\pm_{cond}(\hat A|\hat B)=\min_g \frac{V(\hat A \pm g\hat B)}
{V(\hat A_{SN})}
\end{eqnarray}
being $V(\hat A)$ the variance, and $V(\hat A_{SN})$ the shot noise level.
The factor $g$ is introduced to optimize noise reduction
and is experimentally obtained by an attenuator and a delay line
 \cite{gatti.EPR2}.
We note that
\begin{eqnarray}\nonumber
V(\hat X_1 - \hat X_2 )&<&V(\hat X_{1,SN}) ~~\textrm{and}~~\\
V(\hat Y_1 + \hat Y_2 )&<&V(\hat Y_{1,SN}) \label{EPR.ent.leuch.g=1}
\end{eqnarray}
is a sufficient condition for EPR entanglement, corresponding to
the choice $g=1$. The definition of EPR entanglement used here
\cite{leuch} provides a sufficient condition for the {\em
inseparability} criterion recently discussed for continuous
variables systems in Ref.\cite{Duan}.

For a single-mode type II OPO below threshold, in which transverse
effects are not considered,  EPR correlations  between signal and
idler modes of different polarizations have been predicted and
experimentally demonstrated \cite{reid.opobelow}. Recent
investigations show the possibility of EPR entanglement between
spatial regions of the transverse profile of the signal field of a
degenerate optical parametric oscillator
\cite{gatti.EPR1,gatti.EPR2}. For type II phase matching we can
consider two symmetric far field modes with $x$ and $y$
polarizations, respectively. Neglecting walk-off effects we would
then find
 results equivalent to the degenerate case in type I
phase matching \cite{gatti.EPR1,gatti.EPR2}.
In addition of considering the effect of the walk-off, the novelty of our
discussion here for type II OPO is that
 we can also consider how these correlations change
with the polarization state.

As an indicator to look for EPR entanglement in our case,
we introduce the spectral  variance
of the difference of quadratures
\begin{eqnarray}\label{Eq:Vg.def}
{\mathcal V}_g(\pm\vec k,\omega;\vec \Phi)=&&
\int dt e^{i\omega t}\langle
(\hat A_{\Gamma\Theta}^\Psi(\vec k,t)-g^*\hat A_{\Gamma'\Theta'}^{\Psi'}(-\vec k,t))
\\
\nonumber
&&(\hat A_{\Gamma\Theta}^\Psi(\vec k,0)-g\hat A_{\Gamma'\Theta'}^{\Psi'}(-\vec k,0))
\rangle,
\end{eqnarray}
where the vector $\vec \Phi=(\Gamma,\Theta,\Psi,\Gamma',\Theta',\Psi')$
is the set of parameters determining the polarizations (lower labels)
and quadratures (upper labels) of the symmetric FF modes under consideration.
The value of $g$ giving the minimum variance (\ref{Eq:Vg.def}) is
\begin{eqnarray}\label{Eq:g.def}
\bar{g}=\frac{\int dt e^{i\omega t}\langle
\hat A_{\Gamma'\Theta'}^{\Psi'}(-\vec k,t)
\hat A_{\Gamma\Theta}^\Psi(\vec k,0)\rangle}
{\int dt e^{i\omega t}\langle
\hat A_{\Gamma'\Theta'}^{\Psi'}(-\vec k,t)
\hat A_{\Gamma'\Theta'}^{\Psi'}(-\vec k,0)\rangle}
\end{eqnarray}

From the output moments Eqs. (\ref{A1+A1}-\ref{A1A2}) we obtain
\begin{eqnarray}\label{Eq:Vg}
&&{\mathcal V}_g(\pm\vec k,\omega;\vec \Phi)=\sigma
\\ \nonumber&&\left[
|e^{i(\Psi+\Psi'+\Gamma')}\cos\Theta U_1(\vec k,\omega)-
g^*\sin\Theta' V_2^*(-\vec k,-\omega)|^2 \right. \\ \nonumber
&&+\left. |e^{i(\Psi+\Psi'+\Gamma')}g^* \sin\Theta'  U_1(\vec k,-\omega)-
\cos\Theta V_2^*(-\vec k,\omega)|^2\right. \\  \nonumber
&&+\left. |e^{i(\Psi+\Psi'+\Gamma)} \sin\Theta U_1(-\vec k,-\omega)-
g^*\cos\Theta' V_2^*(\vec k,\omega)|^2\right. \\  \nonumber
&&+\left. |e^{i(\Psi+\Psi'+\Gamma)} g^*\cos\Theta' U_1(-\vec k,\omega)-
\sin\Theta V_2^*(\vec k,-\omega)|^2
\right]
\end{eqnarray}
with shot noise level $\sigma(1+|g|^2)$. A variance below this shot noise
level is a signature of squeezing. We are looking for the more stringent
conditions (\ref{EPR.ent.leuch}) discussed above: EPR entanglement imposes
the requirement that  ${\mathcal V}_g(\pm\vec k,\omega;\vec \Phi)$ goes
below the shot noise level of  $\hat A_{\Gamma\Theta}^\Psi$  (that is we
should find ${\mathcal V}_g(\pm\vec k,\omega;\vec \Phi) <\sigma$),
simultaneously for two combinations of orthogonal quadratures
\cite{leuch}.

The general result (\ref{Eq:Vg})  depends on many
parameters.  However it is important to note that ${\mathcal V}_g$
only depends on the phases $\Psi,\Psi',\Gamma,\Gamma'$ through the
independent combinations $(\Psi+\Psi'+\Gamma)$ and $(\Psi+\Psi'+\Gamma')$.
The dependence on the sum of quadratures angles $\Psi$ and $\Psi'$
is well known in other contexts.
It is easily understood taking into account that measuring by
a single homodyne detector the noise in a quadrature
$\frac{\Psi+\Psi'}{2}$ of the difference of the spatial
$\pm \vec k$ and polarization  ($\Gamma,\Theta,\Gamma',\Theta'$)
modes is equivalent to the noise measurement described by
Eq.(\ref{Eq:Vg.def}) \cite{reid.opobelow}.
The result that ${\mathcal V}_g$ depends on independent variations of
$\Gamma$ and $\Gamma'$ only through their sum with the
 sum of the quadrature phases $\Psi+\Psi'$ means that it is
equivalent to vary the selected quadrature changing the phase of the local
oscillator or to shift both signal and idler fields  by a proper phase
with the phase retarders.

In order to study how  correlations change in different spatial
points of the FF we consider two detection schemes, with detectors in symmetric FF
points, which represent possible extreme cases:
\begin{itemize}
\item In the first detection scheme the field is detected
in the crossing points $\pm\vec k_H$ of the signal and idler rings
in the FF, as represented in Fig. \ref{ch:5.fig:1}. Being these points
in the line $k_y=0$, they are not affected by the transverse walk-off.
In  Fig. \ref{ch:5.fig:1} we indicate the case in which in one point the
polarization component $x$ is selected and in the symmetric point the
polarization $y$ is selected. In Sect. \ref{sect:horizontal} we will
show that for these special points of the FF any change in the selection
of the polarization does not influence the results given by Eq.
(\ref{Eq:Vg}).
\item In the second arrangement, shown  in  Fig. \ref{ch:5.fig:3}, the
detectors are located in a couple of symmetric  points $\pm\vec
k_V$ on the line $k_x=0$, where the walk-off effect is more
pronounced. In Sect. \ref{sect:vertical} we will consider the
effect of changing the polarization state selected. We can
distinguish two extreme cases: Fig.\ref{ch:5.fig:3}a   shows the
arrangement in which the most intense  polarization components on
the rings are detected. We name this scheme as ``vertical bright
scheme". Fig.\ref{ch:5.fig:3}b   shows the case in which the $y$
($x$) polarization component is selected on the intense lower
(upper) $x$ ($y$) polarized ring. We name this scheme as
``vertical dark scheme".
\end{itemize}

\subsection{EPR between far field modes unaffected by walk-off}
\label{sect:horizontal}

In  the line $k_y=0$ there is no  effect of the transverse walk-off and
the coefficients given by  Eqs.(\ref{eq:detunings}),
(\ref{Eq:U1}) and (\ref{Eq:V1}) have the following
reflection symmetry:
\begin{eqnarray}\label{Eq:symmetry}
U_j(k_x,k_y=0)&=&U_j(-k_x,k_y=0)\\ \nonumber
V_j(k_x,k_y=0)&=&V_j(-k_x,k_y=0)
\end{eqnarray}
  with $j=1,2$.
The results presented in this section are strongly dependent on this
symmetry, generally present also in previous treatments of spatial EPR
correlations and squeezing.

Even with this symmetric form of the coefficients the variance
given by Eq.(\ref{Eq:Vg}) is a complicated function of many parameters.
For the sake of simplicity we first consider Eq.(\ref{Eq:Vg}) in
the case of $g=1$  (see Eq.(\ref{EPR.ent.leuch.g=1})). The microscopic
process of generation of twin photons with linear orthogonal polarization
($\Theta=0$ and $\Theta'=\pi/2$)
suggests a natural choice for the relative phases of the polarizers.
Hence we consider a case in which the polarizers
in the symmetric points $\pm \vec k_H$ have a relative phase fixed by
\begin{eqnarray}\label{Eq:phasesTheta}
\Theta'=\Theta+\pi/2.
\end{eqnarray}
Using this relation between $\Theta'$ and $\Theta$, Eq.(\ref{Eq:Vg})
becomes independent of $\Theta$ for $g=1$ when the additional choice
\begin{eqnarray}\label{Eq:phasesGamma}
\Gamma'=\Gamma+\pi
\end{eqnarray}
is also made.
With these particular choice of parameters,
 Eq. (\ref{Eq:Vg}) for the $\pm \vec k_H$ points
(\ref{Eq:kx.crit}) reduces to:
\begin{eqnarray}\nonumber
&&{\mathcal V}_{g=1}(\pm \vec k_H,\omega;[\Gamma,\Theta,\Psi,\Gamma+\pi,
\Theta+\pi/2,\Psi'])= \\\nonumber
&&\sigma \left[~
|e^{i(\Psi+\Psi'+\Gamma)} U_1(\vec k_H,\omega)+V_2^*(\vec k_H,-\omega)|^2
\right.\\
&&+\left. |e^{i(\Psi+\Psi'+\Gamma)}U_1(\vec k_H,-\omega)+
V_2^*(\vec k_H,\omega)|^2\right].
\label{eq:V.horizontal}
\end{eqnarray}
Eq.(\ref{eq:V.horizontal}) explicitly shows that the
fluctuations in any quadrature of the difference of symmetric spatial modes
$\pm \vec k_H$, with relative polarizations fixed by
Eqs.(\ref{Eq:phasesTheta})-(\ref{Eq:phasesGamma}), are independent of the
choice of the polarization reference $\Theta$.

With the above motivation for the relations between phase parameters,
we analyze the EPR correlations in
the $\pm \vec k_H$  points (Fig. \ref{ch:5.fig:1}) fixing the parameters
$\Theta=0$,
$\Gamma=\pi$, $\Theta'=\pi/2$, $\Gamma'=2\pi$,
so that $\hat A_{\Gamma\Theta}^\Psi=\hat A_1^\Psi$
and  $\hat A_{\Gamma'\Theta'}^{\Psi'}=\hat A_2^\Psi$,
and varying the quadrature angles $\Psi$, $\Psi'$.
There is no loss of generality in this choice of $\Gamma$ and $\Theta$
since Eq.(\ref{eq:V.horizontal})
is independent of $\Theta$, and the phase $\Gamma$ can be absorbed in the
quadrature angles $\Psi+\Psi'$.
As discussed previously (see Eq.(\ref{EPR.ent.leuch.g=1})),
EPR entanglement is guaranteed -- for some quadratures ($\Psi+\Psi'$) --
if both the  `position' and `momentum' operators
\begin{eqnarray}\label{Eq:pos.H}
&&\hat A_1^\Psi(\vec k_H)-\hat A_2^{\Psi'}(-\vec k_H)=\\\nonumber
&&[\hat A_1(\vec k_H)e^{i\Psi}+H.c.]-[\hat A_2(-\vec k_H)e^{i\Psi'}+H.c.] \\
\label{Eq:mom.H}
&&\hat A_1^{\Psi+\pi/2}(\vec k_H)+\hat A_2^{\Psi'+\pi/2}(-\vec k_H)=
\\\nonumber
&&[i\hat A_1(\vec k_H)e^{i\Psi}+H.c.]+[i\hat A_2(-\vec k_H)e^{i\Psi'}+H.c.]
\end{eqnarray}
show simultaneously
a variance below the reference value $\sigma$. This value corresponds to
 the shot noise level
of both  $\hat
A_1^\Psi(\vec k_H)$ and $\hat A_1^{\Psi+\pi/2}(\vec k_H)$.
Using Eq. (\ref{Eq:Vg}) it turns out that (\ref{Eq:pos.H}) and
(\ref{Eq:mom.H}) have the same spectral variance. Therefore, in the
following we only
consider Eq. (\ref{Eq:Vg}) normalized to the shot noise level
$\sigma$ for the `position' operator Eq. (\ref{Eq:pos.H}), which is
identified by the angles  $\vec \Phi =(\pi,0,\Psi,0,\pi/2,\Psi')$.

\begin{figure}[t]
\includegraphics[width=0.4\textwidth,clip=true]{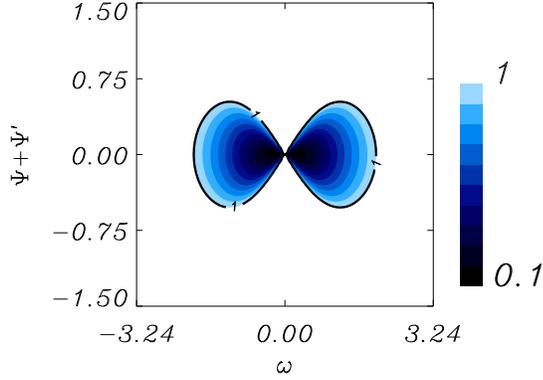}
\caption[${\mathcal V}_{g=1}(\pm \vec k_H,\omega;\vec \Phi)/{\sigma}$]
{${\mathcal V}_{g=1}(\pm \vec k_H,\omega;\vec \Phi)/{\sigma}$
for $\vec \Phi =(\pi,0,\Psi,0,\pi/2,\Psi')$
as a function of $\Psi+\Psi'$ (rad) and the frequency $\omega$.
EPR entanglement is obtained for values less than $1$ (dark line). Values
above $1$ outside this line are not displayed.}
\label{ch:5.fig:hor.2a}
\end{figure}

\begin{figure}[t]
\includegraphics[width=0.4\textwidth,clip=true]{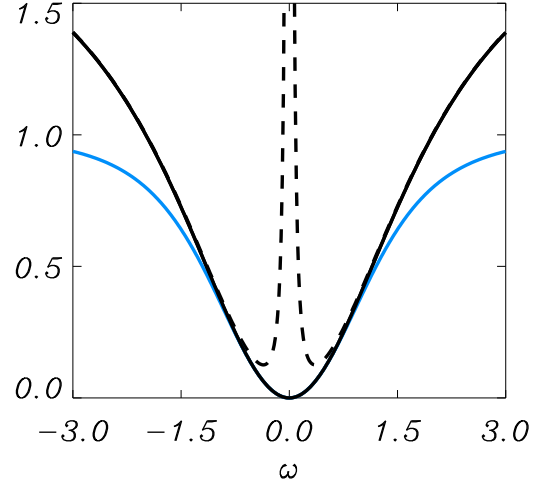}
\caption[Sections of ${\mathcal V}_{g}$]
{${\mathcal V}_{g}(\pm \vec k_H,\omega;\vec \Phi)/
{\sigma}$  for $\Psi+\Psi'=0$ and $g=1$
(dark continuous line),  for $\Psi+\Psi'=0$ and $g=\bar g$
(light continuous line), and for $\Psi+\Psi'\simeq0.06$rad  and $g=1$
(dashed line). }
\label{fig:diff.quadr_hor}
\end{figure}

In Fig. \ref{ch:5.fig:hor.2a} we represent  the normalized variance
${\mathcal V}_{g=1}(\pm \vec k_H,\omega;\vec \Phi)/{\sigma}$ for the points
$\pm \vec k_H$. When this normalized variance
is less than $1$, we find  EPR entanglement. We see a maximum noise
reduction for  $\Psi+\Psi'=0$, and for $\omega=0$.
Fig.\ref{fig:diff.quadr_hor} shows a cut of Fig.\ref{ch:5.fig:hor.2a}
for $\Psi+\Psi'=0$ (continuous
black line).  A variation in the quadrature angle results in a mixing of
squeezed and unsqueezed quadratures degrading the entanglement:
strongest degrading effects are evident at small frequency (dashed black in
Fig.\ref{fig:diff.quadr_hor}).

The fact that there is maximum noise reduction for $\Psi+\Psi'=0$
can be understood by analogy with the result for the single-mode
non-degenerate OPO discussed in Ref.\cite{reid.opobelow}. In fact,
for the case of vanishing detunings of the signal and idler and
for real pump, which was the one considered in Ref.
\cite{reid.opobelow}, the squeezing direction corresponds to the
real quadratures ($\Psi=\Psi'=0$). In our case we find an
analogous result
 because  we are also considering a real pump, and
the effective detunings, given by (\ref{eq:detunings}),
also vanish: $\Delta_j(\vec k=\pm\vec k_H,\omega=0)=0$.

Our calculations allows us to search for EPR entanglement considering any
couple of symmetric FF modes (distinguished by their transversal
wavevector) and  selecting \emph{any} polarization.   We have found EPR
entanglement for the modes $\pm \vec k_H$ for any  choice of the polarization
component in $- \vec k_H$ (varying $\Theta$), if in $+ \vec k_H$ we select the
orthogonal polarization ($\Theta'=\Theta+\pi/2$).  In these far field
points $\pm \vec k_H$, which are not affected by the walk-off,  any mixing of
the the signal and idler fields detected in a point is entangled with the
field in the symmetric point, if this is also properly mixed
(Eqs.(\ref{Eq:phasesTheta})-(\ref{Eq:phasesGamma})).

\begin{figure}
\includegraphics[width=0.4\textwidth,clip=true]{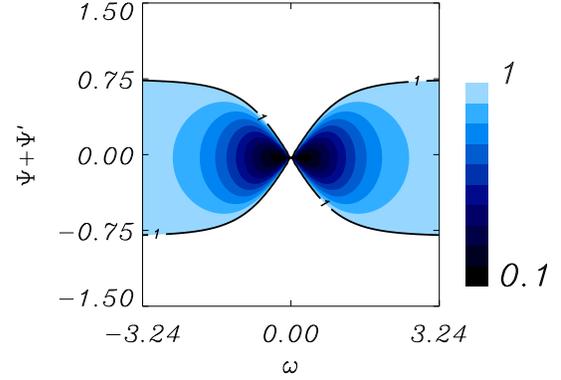}
\caption[${\mathcal V}_{\bar{g}}(\pm \vec k_H,\omega;\vec \Phi)/{\sigma}$]{
Same as in Fig.\ref{ch:5.fig:hor.2a}, but for the choice
 $g=\bar{g}$.}
\label{ch:5.fig:hor.g.2a}
\end{figure}

So far we have considered the case $g=1$ and we have found that a
sufficient condition to  guarantee EPR entanglement between the
modes $\pm \vec k_H$ is fulfilled (see
Eq.\ref{EPR.ent.leuch.g=1}). If we now optimize the noise
reduction considering  $g=\bar g$ we obtain the results shown in
Fig.\ref{ch:5.fig:hor.g.2a}.  Comparing the normalized variance
${\mathcal V}_{g=1}$ (Fig. \ref{ch:5.fig:hor.2a}) with  ${\mathcal
V}_{\bar g}$ (Fig. \ref{ch:5.fig:hor.g.2a}) we observe that with
the choice  $g=\bar g$,  EPR entanglement is observed in a larger
frequency bandwidth. For a more direct comparison of the variances
obtained for $g=\bar g$ and  $g=1$, both quantities are plotted
for $\Psi+\Psi'=0$ in Fig.\ref{fig:diff.quadr_hor}: for small
frequencies the results are very similar, while for increasing
values of the frequencies  the choice  $g=\bar g$ allows to
observe EPR entanglement even when this effect is lost for  $g=1$.

Our strong EPR correlations have been obtained for
$\Theta'=\Theta+\pi/2$.  As already
mentioned, this phase relation corresponds to the underlying process governing
 the creation of twin photons with orthogonal
polarizations. In addition, for $g=1$ we have obtained that the
EPR correlations are independent of $\Theta$. If we decrease the
relative angle between $\Theta$ and $\Theta'$, we observe a
progressive reduction of the entanglement. In the limiting case
$\Theta=\Theta'$ the correlation between symmetric spatial modes
is vanishing and ${\mathcal V}_{g}>\sigma$. In fact in this case
no twin photons are detected.

\subsection{EPR between far field modes in the walk-off direction}
\label{sect:vertical}

\begin{figure}[t]
\includegraphics[width=0.4\textwidth,clip=true]{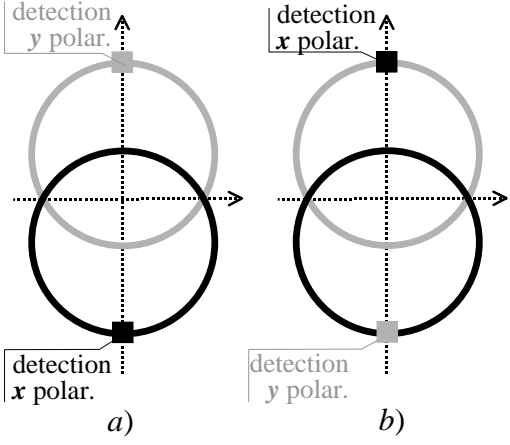}
\caption[Vertical detection schemes]{FF signal and idler rings and
detection scheme influenced by the walk-off: a) The $x$ and $y$ polarized
fields are detected in the points $\pm\vec k_V$ indicated by the square
symbols, on the intense  $x$ polarized (black) and $y$ polarized (grey)
rings  (bright detection). b) In the same points $\pm\vec k_V$ (square
symbols)  the orthogonal  polarizations $y$ and $x$ are detected (dark
detection). }
\label{ch:5.fig:3}
\end{figure}

In this section we study   possible EPR entanglement  for
symmetric points of the far field along the  walk-off  direction
($y$), in the arrangements shown in Fig.\ref{ch:5.fig:3}. An
important effect of the transverse walk-off is   breaking the
reflection symmetry in the far field. This symmetry is generally
broken for $k_y\neq 0$. For the points shown in
Fig.\ref{ch:5.fig:3}
\begin{eqnarray}
U_j(k_x=0,k_y)&\neq& U_j(k_x=0,-k_y)~~~~~~~~~j=1,2\\
V_j(k_x=0,k_y)&\neq& V_j(k_x=0,-k_y)~~~~~~~~~j=1,2.
\end{eqnarray}

Following the considerations in the previous Section,
 we will also consider here the case of phase polarizations
$\Theta,\Theta',\Gamma,\Gamma'$ fixed by
Eqs.(\ref{Eq:phasesTheta})-(\ref{Eq:phasesGamma}). For this special choice of
phase relations, the variance
 (\ref{Eq:Vg}) for arbitrary $g$ reduces to:
\begin{eqnarray}\label{Eq:V.vertical}
&&{\mathcal V}_g(\pm\vec k,\omega;\vec \Phi)=\\  \nonumber
&&\sigma \{  \cos^2\Theta
[|e^{i(\Psi+\Psi'+\Gamma)} U_1(\vec k,\omega)+
g^*V_2^*(-\vec k,-\omega)|^2\\  \nonumber
&&+|e^{i(\Psi+\Psi'+\Gamma)}g^*  U_1(\vec k,-\omega)+
V_2^*(-\vec k,\omega)|^2 ]\\  \nonumber
&&+  \sin^2\Theta
[|e^{i(\Psi+\Psi'+\Gamma)} U_1(-\vec k,-\omega)+
g^*V_2^*(\vec k,\omega)|^2\\  \nonumber
&&+|e^{i(\Psi+\Psi'+\Gamma)}g^*  U_1(-\vec k,\omega)+
V_2^*(\vec k,-\omega)|^2 ]
\}   .
\end{eqnarray}
The lack of  reflection  symmetry  implies that the variance
${\mathcal V}_g(\pm\vec
k,\omega;\vec \Phi)$  depends now on the angle $\Theta$ and a simple result
analogous to  Eq.(\ref{eq:V.horizontal}) cannot be obtained when
$k_y\neq 0$.
Eq.(\ref{Eq:V.vertical}) depends on  the sum
$\Psi+\Psi'+\Gamma$: therefore, without loss of generality, we can absorb
the effect of the wave retarder in the phase of the local oscillator,
fixing  the angle  $\Gamma=\pi$ as in  Sect.\ref{sect:horizontal}. In
the following we consider  the dependence of Eq.(\ref{Eq:V.vertical}) on the
angle $\Theta$. Varying $\Theta$ different polarization components are
selected locally. We will see that the selection of different values of
$\Theta$ can improve or degrade  EPR entanglement.
In particular  we consider two values for
the angle $\Theta$ ($=0,\pi/2$) leading to very different situations.

\begin{figure}[t]
\includegraphics[width=0.4\textwidth,clip=true]{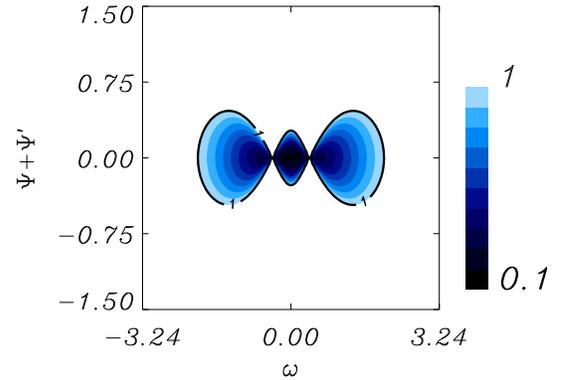}
\caption[${\mathcal V}_{g=1}(\pm \vec k_V,\omega;\vec \Phi)/{\sigma}$]
{${\mathcal V}_{g=1}(\pm\vec  k_V,\omega;\vec \Phi)/{\sigma}$
for $\vec \Phi =(\pi,\pi/2,\Psi,0,\pi,\Psi')$
as a function of $\Psi+\Psi'$ (rad) and the frequency $\omega$.
The detection scheme is the one shown in Fig.\ref{ch:5.fig:3}a.
EPR entanglement is obtained for values less than $1$ (dark line). Values
above $1$ outside this line are not displayed.}
\label{ch:5.fig:vgood.2a}
\end{figure}

\begin{figure}[t]
\includegraphics[width=0.4\textwidth,clip=true]{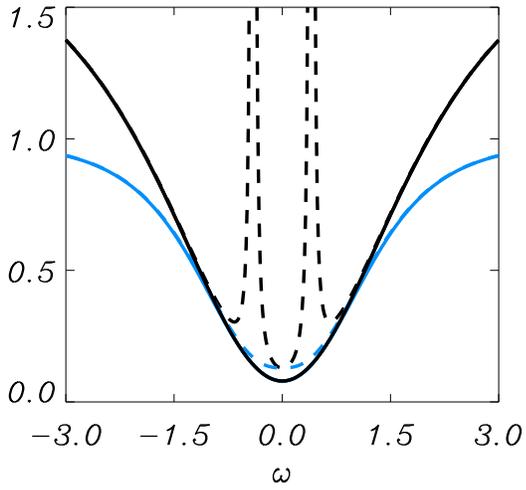}
\caption[Sections of ${\mathcal V}_{g}$]
{Dependence of ${\mathcal V}_{g}(\pm \vec k_V,\omega;\vec \Phi)/
{\sigma}$ on the frequency $\omega$,  for $\Psi+\Psi'=0$ and $g=1$
(dark continuous line),  for $\Psi+\Psi'=0$ and $g=\bar g$
(light continuous line),  for $\Psi+\Psi'\simeq0.06$rad  and $g=1$
(dark dashed line), and for $\Psi+\Psi'\simeq0.06$rad  and $g= \bar g$
(light dashed line).}
\label{fig:diff.quadr_vgood}
\end{figure}

First we consider the case represented in Fig.\ref{ch:5.fig:3}a,
in which the polarizer is oriented so that the most intense linear
polarization component is selected locally. We are selecting two
critical spatial modes   $\hat A_1(-\vec k_{V})$ and $\hat
A_2(\vec k_{V})$,  whose quantum fluctuations  are weakly damped.
This is the  ``vertical bright" detection scheme. In order to
detect the intense polarization components at $\pm \vec k_{V}$ the
phase $\Theta$  must be fixed at
\begin{eqnarray}
\Theta=\pi/2
\end{eqnarray}
and, given Eq.(\ref{Eq:phasesTheta}), $\Theta'=\pi$. Therefore the phases
to be
considered are $\vec \Phi =(\pi,\pi/2,\Psi,0,\pi,\Psi')$.
We first consider the case $g=1$. We look for  EPR entanglement
between the  `position' and `momentum' quadratures
\begin{eqnarray}\label{Eq:pos.mom.Vgood}
\hat A_1^{\Psi'}(-\vec k_{V})-\hat A_2^\Psi(\vec k_{V}), ~
\hat A_1^{\Psi'+\pi/2}(-\vec k_{V})+\hat A_2^{\Psi+\pi/2}(\vec k_{V}).
\end{eqnarray}
We will only show the variance of the `position' quadrature, the
orthogonal one being equivalent. We obtain EPR entanglement, as
shown in Fig.\ref{ch:5.fig:vgood.2a}.  Maximum noise reduction is
obtained for  $\Psi+\Psi'=0$. The variance ${\mathcal V}_{g=1}$
normalized to $\sigma$ and for  $\Psi+\Psi'=0$ is represented as a
function of the frequency $\omega$ in
Fig.\ref{fig:diff.quadr_vgood}: the best entanglement is observed
for $\omega=0$.  Also in this case a variation in the quadrature
angle $\Psi+\Psi'$ results in a mixing of squeezed and unsqueezed
quadratures degrading the entanglement
(Fig.\ref{fig:diff.quadr_vgood}). We note an important difference
with respect to the case for $\pm \vec k_H$: in
Fig.\ref{fig:diff.quadr_hor} the largest degradation of
entanglement for $\Psi+\Psi'\neq 0$ was observed for vanishing
frequency ($\omega\simeq 0$), while in
Fig.\ref{fig:diff.quadr_vgood} we see that  for vanishing
frequency the entanglement  is only partially degraded.  The
largest degradation occurs now for  $\omega\simeq \pm 0.4045$ (two
peaks in Fig.\ref{fig:diff.quadr_vgood}). This value of frequency
coincides with the Hopf frequency $\omega_H$ for the mode $\vec
k_V$, as can be easily checked by Eq.(\ref{Eq:wH}).

\begin{figure}
\includegraphics[width=0.4\textwidth,clip=true]{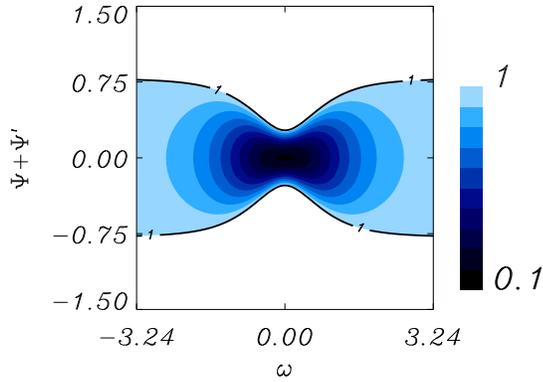}
\caption[${\mathcal V}_{g=\bar{g}}(\pm \vec k_V,\omega;\vec \Phi)/{\sigma}$]
{${\mathcal V}_{g=\bar{g}}(\pm\vec  k_V,\omega;\vec \Phi)/{\sigma}$
as in Fig.\ref{ch:5.fig:vgood.2a}, but for the choice
 $g=\bar{g}$.}
\label{ch:5.fig:vgood.g.2a}
\end{figure}

In Fig. \ref{ch:5.fig:vgood.g.2a} we show the optimized
($g=\bar{g}$) variance ${\mathcal V}_{g=\bar{g}}$
(\ref{Eq:V.vertical}) normalized to the shot noise $\sigma$.  We
obtain EPR maximum entanglement  for $\Psi+\Psi'=0$ and for small
frequencies. Both the minimum and maximum fluctuations are
obtained in a bandwidth of frequencies centered in zero, as in the
case for the points $\pm \vec k_H$. The effects of the Hopf
frequency found for $g=1$ disappear for $g=\bar{g}$. A clear
difference with respect to the variance ${\mathcal V}_{g=\bar{g}}$
obtained in Sect.\ref{sect:horizontal} is the level of maximum
noise suppression reached. Even if in both cases we observe strong
EPR entanglement, in this "vertical bright" arrangement  the
quadratures correlations are reduced, as can be seen comparing
Fig.\ref{fig:diff.quadr_hor} and Fig.\ref{fig:diff.quadr_vgood}.
This reduction is caused by the walk-off.  The fields in the
critical modes $\pm \vec k_{H}$ --not affected by  walk-off-- have
vanishing effective detunings  (\ref{eq:detunings}) for the
threshold frequency $\omega_H(\vec k_{H})= 0$; while  the fields
in  the critical modes $\pm\vec k_{V}$ --in the walk-off
direction-- have vanishing effective detunings
(\ref{eq:detunings}) for the threshold çfrequency
$\omega_H(\pm\vec k_{V})\neq 0$. The fact that the detunings do
not vanish for  $\omega= 0$ seems to be the mechanism responsible
for the reduction of squeezing at $\omega= 0$ for the modes
$\pm\vec k_{V}$. On the other hand, also the unsqueezed quadrature
is influenced by this effect, showing a reduced amplification with
respect to the values obtained for the points $\pm \vec k_H$.
Comparing   Figs.\ref{ch:5.fig:hor.g.2a} and
\ref{ch:5.fig:vgood.g.2a},  we see that in this  "vertical bright"
arrangement the variance is less sensible to deviations of the
quadrature phases selected, with respect to the optimum choice
$\Psi+\Psi'=0$. In Fig.\ref{ch:5.fig:vgood.g.2a} we observe a
broad interval of phases $\Psi+\Psi'$ giving EPR entanglement for
$\omega= 0$.

\begin{figure}
\includegraphics[width=0.4\textwidth,clip=true]{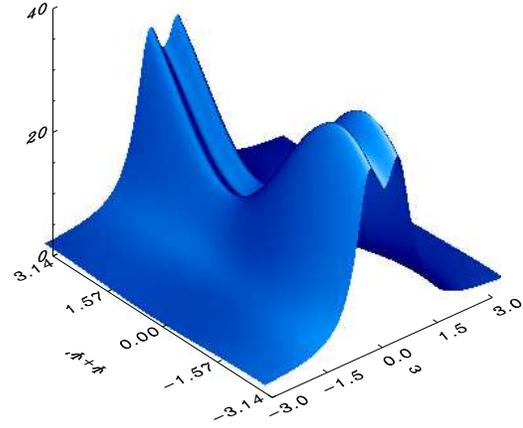}
\caption[${\mathcal V}_{g=1}(\pm \vec k_V,\omega;\vec \Phi)/{\sigma}$]
{${\mathcal V}_{g=1}(\pm\vec  k_V,\omega;\vec \Phi)/{\sigma}$
for  $\vec \Phi =(\pi,0,\Psi,0,\pi/2,\Psi')$,
as a function of $\Psi+\Psi'$ (rad) and frequency $\omega$.
The detection scheme is the one shown in Fig.\ref{ch:5.fig:3}b.}
\label{ch:5.fig:vbad.2a}
\end{figure}

\begin{figure}
\includegraphics[width=0.4\textwidth,clip=true]{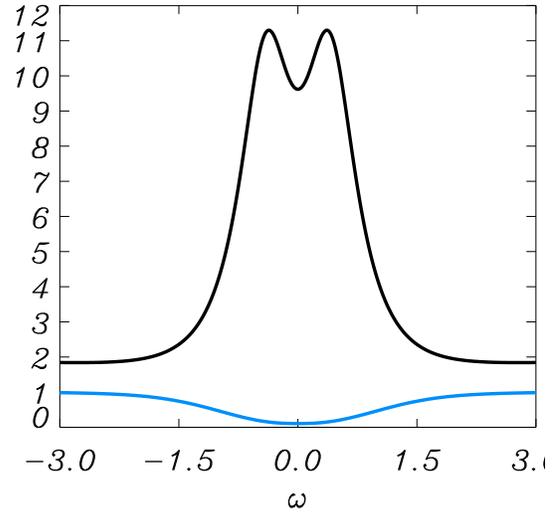}
\caption[Sections of ${\mathcal V}_{g}$]
{Dependence of ${\mathcal V}_{g}(\pm \vec k_V,\omega;\vec \Phi)/
{\sigma}$ on the frequency $\omega$,  for $\Psi+\Psi'=\bar\Psi
$ and $g=1$
(dark continuous line),  for $\Psi+\Psi'=\bar\Psi$ and $g=\bar g$
(light continuous line). $\bar\Psi\simeq0.81$rad.}
\label{fig:diff.quadr_vbad}
\end{figure}

Next we consider  the detection scheme of Fig.\ref{ch:5.fig:3}b.
In this case the phase of the polarizer at $+\vec k_{V}$ is fixed at
\begin{eqnarray}
\Theta=0,
\end{eqnarray}
and $\vec \Phi =(\pi,0,\Psi,0,\pi/2,\Psi')$. With this selection
of the phases of the polarizers  the intense field component  are
filtered out. In this ``vertical dark" detection scheme, the
detected modes ($\hat A_1(\vec k_{V})$ and $\hat A_2(-\vec
k_{V})$) have low intensities. The main point is that now we are
considering non critical modes that are strongly damped at any
frequency (see Sect.\ref{sect:1}). We evaluate again the spectral
variances of the position and momentum quadratures. In this case
the  results obtained for $g=1$ and $g=\bar{g}$ are completely
different. We start considering  Eq.(\ref{Eq:V.vertical}) for
$g=1$. We obtain that ${\mathcal V}_{g=1}$ is always larger than
$\sigma$, therefore no EPR entanglement is observed  (see
Fig.\ref{ch:5.fig:vbad.2a}).  In the same way that in  the
vertical bright scheme,  the largest fluctuations are observed at
$\omega\neq0$. Fig.\ref{fig:diff.quadr_vbad} shows
 a cut of  Fig.\ref{ch:5.fig:vbad.2a} for the quadrature
$\bar\Psi$, for which there is a minimum in the direction $\Psi+\Psi'$.

\begin{figure}
\includegraphics[width=0.4\textwidth,clip=true]{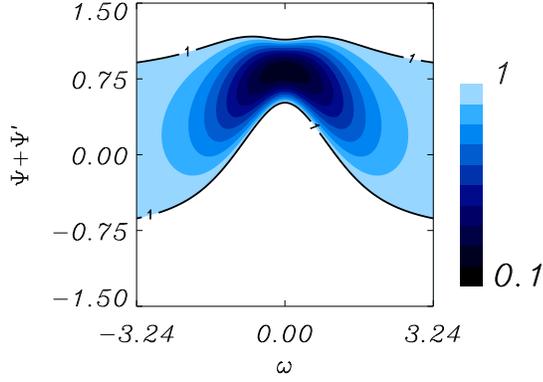}
\caption[${\mathcal V}_{g=\bar{g}}(\pm \vec k_V,\omega;\vec \Phi)/{\sigma}$]
{Same as in Fig.\ref{ch:5.fig:vbad.2a}, but for the choice of
 $g=\bar{g}$.EPR entanglement is obtained for values less than $1$ (dark line). Values
above $1$ outside this line are not displayed.}
\label{ch:5.fig:vbad.g.2a}
\end{figure}

We now consider the variance (\ref{Eq:V.vertical}) for the best
choice  $g=\bar{g}$. We find that  ${\mathcal V}_{\bar{g}}$, represented
in Fig.\ref{ch:5.fig:vbad.g.2a}, is reduced below the shot noise level
$\sigma$ for a large region of parameters.
Therefore, with a proper choice of $g$, EPR entanglement is obtained
also in this case. From Fig.\ref{ch:5.fig:vbad.g.2a} we also see that for
$\Psi+\Psi'=0$ only  small entanglement would be
observed, and in the region of large frequencies. In fact,
the quadrature at which strong EPR effects are
observed is $\Psi+\Psi'=\bar\Psi\neq 0$.
Fig.\ref{fig:diff.quadr_vbad} shows  ${\mathcal V}_{\bar{g}}$ for this choice
of $\Psi+\Psi'$. Changing the walk-off parameter we have found that
$\bar\Psi$ increases with the walk-off.

\begin{figure}[t]
\includegraphics[width=0.4\textwidth,clip=true]{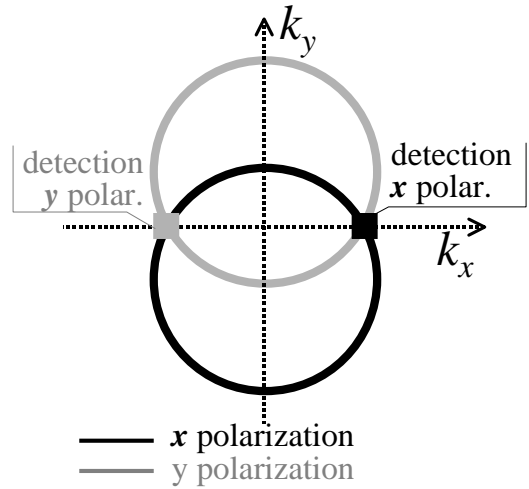}
\caption[Horizontal detection scheme]{FF signal (black) and idler
(grey) rings and detection scheme not influenced by the walk-off:
The $x$ and $y$ polarizations are detected at the points $\pm\vec
k_H$ where the rings intersect (square symbols).}
\label{ch:5.fig:1}
\end{figure}

\begin{figure}
\includegraphics[width=0.4\textwidth,clip=true]{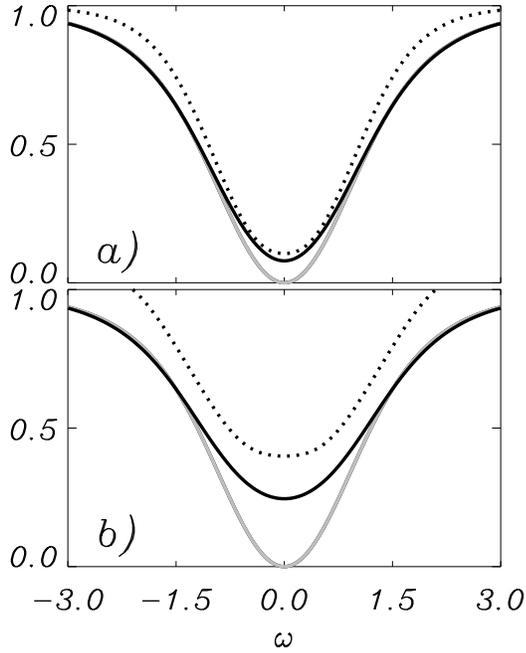}
\caption[${\mathcal V}_{\bar{g}}$ in different detection schemes]{Spectral variance ${\mathcal V}_{\bar{g}}$
as a function of  frequency, for the detection schemes represented in
Fig.\ref{ch:5.fig:1} (light line),  Fig.\ref{ch:5.fig:3}a (dark continuous
line), and  Fig.\ref{ch:5.fig:3}b (dark dotted line).
The angles $\Psi+\Psi'=0,0,0.81$ rad are selected respectively for each
detection scheme.
a) walk-off $\rho_2=1$;
b) walk-off $\rho_2=1.5$. } \label{ch:5.fig:6}
\end{figure}

In Fig.\ref{ch:5.fig:6} we show a comparison of the best EPR entanglement
($g=\bar g$, optimum $\Psi+\Psi'$) found for the three detection schemes
considered, namely Fig.\ref{ch:5.fig:1},  Fig.\ref{ch:5.fig:3}a and
Fig.\ref{ch:5.fig:3}b.
We observe that  the  correlations are less important when we move
 from the  detection scheme of Fig.\ref{ch:5.fig:1} to the
the ``vertical bright" and ``vertical dark" schemes of
Fig.\ref{ch:5.fig:3}a and Fig.\ref{ch:5.fig:3}b. We also show the
effect of increasing the walk-off: we can see that EPR
entanglement in the ``vertical dark" and  ``bright" schemes  get
worst  increasing the walk-off strength. Obviously the results for
the detection scheme of Fig. \ref{ch:5.fig:1} are not influenced
by walk-off.


\section{Stokes operators}
\label{sect:3}
In the previous Section we have seen how the selection of different
polarization components in the far field influences the
quadratures EPR entanglement
between symmetric FF points. The  results we have obtained also
show the effects of
the transverse walk-off.
In this Section our aim is to characterize the polarization properties
of a type II OPO, when  transverse walk-off is taken into account.
The polarization state of the  field in any point of the transverse plane
can be characterized in terms of the Stokes parameters
\cite{born-wolf}.
 An operational definition
of the Stokes parameters can be given using polarizers and
retarders \cite{born-wolf,hecht}. In the quantum formalism there
are different ways to describe the polarization giving the same
classical limit \cite{Hakioglu,Tsegaye}. Here we consider the
quantum Stokes operators $\hat{S}_j(\vec{k},t)$ ($j=0,1,2,3$), for
each FF mode $\vec{k}$, obtained replacing   by creation and
annihilation operators the corresponding observables in the
classical definitions (see Ref. \cite{Usachev} and references
therein):
\begin{eqnarray}
\hat{S}_0(\vec{k},t)&=&\hat{A}^\dagger_1(\vec{k},t)\hat{A}_1(\vec{k},t)
+\hat{A}^\dagger_2(\vec{k},t)\hat{A}_2(\vec{k},t) \label{eq.S0}
\end{eqnarray}
is the total intensity operator,
\begin{eqnarray}
\hat{S}_1(\vec{k},t)&=&\hat{A}^\dagger_1(\vec{k},t)\hat{A}_1(\vec{k},t)
-\hat{A}^\dagger_2(\vec{k},t)\hat{A}_2(\vec{k},t) \label{eq.S1}
\end{eqnarray}
gives the difference  between the $x$ and $y$ linear polarizations,
\begin{eqnarray}
\hat{S}_2(\vec{k},t)&=&\hat{A}^\dagger_1(\vec{k},t)\hat{A}_2(\vec{k},t)
+\hat{A}^\dagger_2(\vec{k},t)\hat{A}_1(\vec{k},t) \label{eq.S2}
\end{eqnarray}
gives the difference  between the $45^\circ$ and $135^\circ$  linear
polarizations, and
\begin{eqnarray}
\hat{S}_3(\vec{k},t)&=&-i(\hat{A}^\dagger_1(\vec{k},t)\hat{A}_2(\vec{k},t)
-\hat{A}^\dagger_2(\vec{k},t)\hat{A}_1(\vec{k},t)) \label{eq.S3}
\end{eqnarray}
gives the difference   between the right-handed and the left
handed circular  polarizations components \cite{nota}.
The definitions (\ref{eq.S0})-(\ref{eq.S3}) correspond, except for a
constant, to the the Schwinger transformation of the modes
$\hat{A}_1(\vec{k},t)$ and $\hat{A}_2(\vec{k},t)$ giving
operators satisfying  angular momentum commutation relations
 \cite{Soderholm}:
\begin{eqnarray}
\left[ \hat{S}_0(\vec{k},t),\hat{S}_j(\vec{k}',t') \right]&=&0 \\
\left[ \hat{S}_j(\vec{k},t),\hat{S}_k(\vec{k}',t') \right]&=&
2i\epsilon_{jkl}
\hat{S}_l(\vec{k},t)\delta(t-t')\sigma\delta_{\vec{k}\vec{k}'}
\end{eqnarray}
with $j,k,l=1,2,3$ \cite{notabis}.
The precision of simultaneous measurements of
the Stokes operators is limited by the Heisenberg principle. For instance
we have
\begin{eqnarray}\label{eq:heis}
\Delta^2\hat{S}_1(\vec{k},t)\Delta^2\hat{S}_2(\vec{k},t')
&\geq& \frac{1}{4} |\langle[\hat{S}_1(\vec{k},t),
\hat{S}_2(\vec{k},t')]\rangle|^2\\\nonumber
&=&|\langle\hat{S}_3(\vec{k},t)\rangle|^2\delta(t-t').
\end{eqnarray}
The Stokes vector  $\hat{\bf{S}}=(\hat{S}_1,\hat{S}_2,\hat{S}_3)$
can be represented in a quantum Poincar\'e sphere, with a  radius
defined by $\langle \hat{S}_0\rangle $  (see for instance
Ref.\cite{Korolkova.2002}). Given the fluctuations of $\hat{S}_i$,
the quantum states are not defined by points on the surface of
this sphere, but rather they are defined by different volumes,
such  as spheres (coherent states) or ellipsoids (squeezed
states).  These quantum uncertainty volumes on the Poincar\'e
sphere have been confirmed by recent experiments
\cite{Bachor.2002}. The transformations Eqs.(\ref{eq:W}) and
(\ref{eq:P}) introduced in the previous Section can be also
visualized in the Poincar\'e sphere. In fact they correspond to a
rotation in the Poincar\'e Sphere of an angle $2\Theta$ around the
$S_3$ axis and of $-\Gamma$ around the $S_1$ axis.

\subsection{Far field local properties}
\label{sect:3.local} 
\begin{figure}
\includegraphics[width=0.4\textwidth,clip=true]{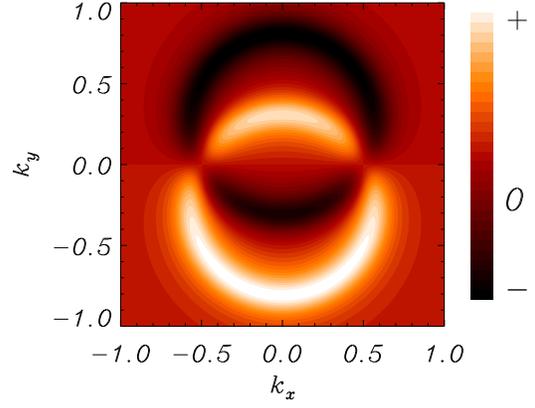}
\caption[$\langle\hat{S}_1(\vec{k})\rangle$]{Stationary average of the Stokes operator
$\langle\hat{S}_1(\vec{k})\rangle$ in the FF (Eq.(\ref{eq:s2.average})).}
\label{ch:5.fig:7}
\end{figure}

Given Eqs. (\ref{A1+A1}-\ref{A1A2}) we obtain
the stationary value of the average of the Stokes parameters. The
average of  $\hat{S}_0$ is given in Eq. (\ref{Eq:It}) and for the
others Stokes parameters we find:
\begin{eqnarray}\label{eq:s2.average}
\langle\hat{S}_1(\vec{k},t)\rangle&=&
\frac{\sigma}{2\pi}\int
d\omega (|V_1(\vec k,\omega)|^2-|V_2(\vec k,\omega)|^2)\\
\label{eq:s3.average}
\langle\hat{S}_2(\vec{k},t)\rangle&=&
\langle\hat{S}_3(\vec{k},t)\rangle=0.
\end{eqnarray}
The average of $\hat{S}_2$ and $\hat{S}_3$ vanishes in any point of the
far field, because any signal or idler photon has the same probability to
be measured along the $45^\circ$ and $135^\circ$ polarizations directions.
The same is true for the left and right circular polarizations. The equal
average intensities at the output of the beam splitters are subtracted,
giving vanishing values of $\langle\hat{S}_2\rangle$ and
$\langle\hat{S}_3\rangle$ independently of the relative intensity of the
signal and the idler.
In Fig. \ref{ch:5.fig:7} we show the far field
spatial profile of  $\langle\hat{S}_1\rangle$:  the lower ring is
dominated by the linear polarization $x$ while the upper one is dominated
by the $y$ polarization.  If there was no walk-off, $\langle\hat{S}_1\rangle$
would vanish in all the FF, while in our case
($\rho_1,\rho_2\neq0$) it only  vanishes along the direction $k_y=0$.
Therefore, in the $\pm\vec k_H$ points we have an intense field (see
$\hat{S}_0$ in Fig. \ref{ch:5.fig:0}) with vanishing average of the Stokes
vector $\hat{\bf{S}}$ and with variances of $\hat{S}_i$
$not$ limited by the Heisenberg
principle, since from  Eq. (\ref{eq:heis})
$|\langle[\hat{S}_i(\vec{k},t),\hat{S}_j(\vec{k}',t')]\rangle|=0$, $i\neq j$.

\begin{figure}
\includegraphics[width=0.4\textwidth,clip=true]{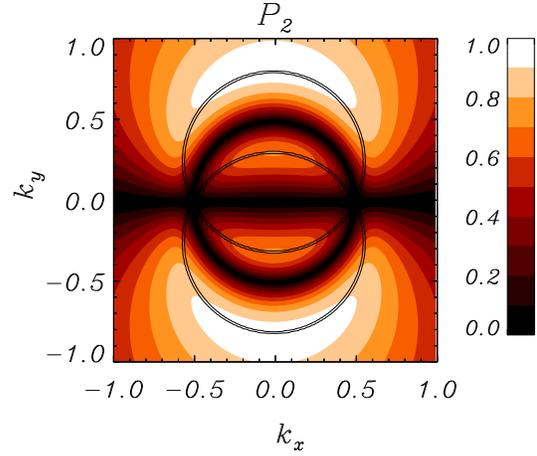}
\caption[$P_2(\vec k,t)$]{Second order polarization degree $P_2(\vec k,t)$
defined in Eq.
(\ref{eq:pol.degree}). The two circles of double continuous line
show the signal and idler maxima intensities.}
\label{ch:5.fig:8}
\end{figure}

The parameter that corresponds to the classical characterization
of  the polarization state of a
quasi-monochromatic field, is the second order  polarization degree
\begin{eqnarray}\label{eq:pol.degree}
P_2(\vec k,t)=\frac{\sqrt{\sum_{j=1}^3\langle\hat{S}_j(\vec k,t)\rangle^2}}
{\langle\hat{S}_0(\vec k,t)\rangle}
\end{eqnarray}
varying from $P_2=0$ for unpolarized light, to $P_2=1$ for completely
polarized light.
In Fig. \ref{ch:5.fig:8} we observe how the  polarization degree, that
reduces to
\begin{eqnarray}
P_2=\frac{|\langle\hat{S}_1\rangle|}
{\langle\hat{S}_0\rangle},
\end{eqnarray}
varies in the FF: In particular, the intense FF rings are always
polarized except around  the line $k_y=0$, where $P_2$ vanishes.
Therefore, the field in the  points $\pm\vec k_H$ is unpolarized
in the ordinary sense. However the concept of polarized and
unpolarized light needs to be generalized in quantum optics
\cite{Klyshko-97,Usachev,Soderholm,Prakash,Luis}. The fact that
$\langle\hat{S}_j\rangle=0$ ($j=1,2,3$), so that $P_2$ vanishes,
does not guarantee to have an unpolarized state from a quantum
point of view. Rather one has to consider the values of the higher
order input moments of $\hat{S}_j$. In Ref.\cite{Klyshko-97} it
was shown that in the single-mode type II PDC, the squeezed vacuum
is unpolarized in the ordinary sense ($P_2=0$) due to the
diffusion in the difference of the signal and idler phases. On the
other hand, due to the twin photon creation, there is complete
noise suppression in the intensity difference of the two linearly
polarized modes, i.e. in $\hat{S}_1$, leading to polarization
squeezing \cite{Chirkin-93}. Due to this anisotropic distribution
of the fluctuations in the Stokes vector $\hat{\bf{S}}$  there is
a \emph{``hidden polarization"} \cite{Klyshko-97},  which has been
observed experimentally recently \cite{Usachev}. We note that
similar hidden polarization should be observed locally in the
transverse near field plane of a type II OPO. Our interest here is
in the far field plane, where twin photons are spatially
separated.

In order to characterize the far field polarization properties, we proceed to
evaluate the variance of the Stokes parameters.
We define the spectral correlation function
\begin{eqnarray}\label{eq:gammai}
&&\Gamma_{i}(\vec{k},\vec{k}',\Omega)=\\\nonumber
&&\int dt e^{i\Omega t}
[\langle\hat{S}_i(\vec{k},t)\hat{S}_i(\vec{k}',0)\rangle-
\langle\hat{S}_i(\vec{k},t)\rangle
\langle\hat{S}_i(\vec{k}',0)\rangle].
\end{eqnarray}
Given the moments (\ref{A1+A1}-\ref{A1A2}) and using the moments
theorem \cite{Goodman},
we obtain non vanishing contributions only for $\vec{k}'=\vec{k}$ (self
correlation) and for $\vec{k}'=-\vec{k}$ (twin photons correlation).
For $\Omega=0$, corresponding to
integrating over a time interval long enough
with respect to the cavity lifetime, the self
correlation of $\hat{S}_1$ is
\begin{eqnarray}\label{eq:gamma1.self}
&&\Gamma_{1}(\vec{k},\vec{k},0)= \\\nonumber
&&\sigma^2\int \frac{d\omega}{2\pi}
[|U_1(\vec{k},\omega)|^2|V_1(\vec{k},\omega)|^2+
 |U_2(\vec{k},\omega)|^2|V_2(\vec{k},\omega)|^2]
\end{eqnarray}
while the twin photons correlation for symmetric FF points is
\begin{eqnarray}\label{eq:gamma1.twin}
\Gamma_{1}(\vec{k},-\vec{k},0)= -\Gamma_{1}(\vec{k},\vec{k},0).
\end{eqnarray}
It can be easily shown that
the second and the third Stokes operators have equivalent variances:
\begin{eqnarray}
\Gamma_{2}(\vec{k},\vec{k}',\omega)=\Gamma_{3}(\vec{k},\vec{k}',\omega).
\end{eqnarray}
For the self correlations we obtain
\begin{eqnarray}\label{eq:gamma2.self}
&&\Gamma_{2,3}(\vec{k},\vec{k},0)= \\\nonumber
&&\sigma^2\int \frac{d\omega}{2\pi}
[|U_2(\vec{k},\omega)|^2|V_1(\vec{k},\omega)|^2+
 |U_1(\vec{k},\omega)|^2|V_2(\vec{k},\omega)|^2]
\end{eqnarray}
and for the twin correlations
\begin{eqnarray}\label{eq:gamma2.twin}
&&\Gamma_{2,3}(\vec{k},-\vec{k},0)= \\\nonumber
&&\sigma^2\int \frac{d\omega}{2\pi}
[U_1^*(\vec{k},\omega)U_1(-\vec{k},-\omega)
 V_2^*(-\vec{k},-\omega)V_2(\vec{k},\omega)+c.c.]
\end{eqnarray}

\begin{figure}
\includegraphics[width=0.5\textwidth,clip=true]{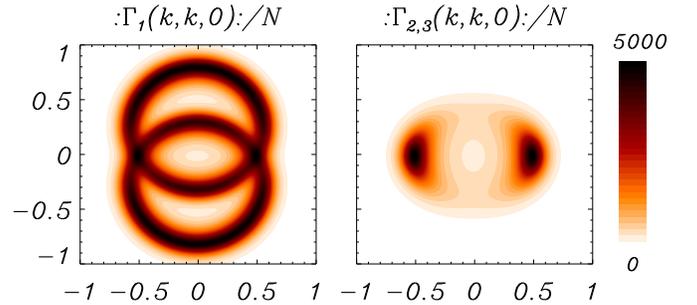}
\caption[$\Gamma_{1}(\vec{k},\vec{k},0)$ and
$\Gamma_{2,3}(\vec{k},\vec{k},0)$]
{Normal ordered variances
$\Gamma_{1}(\vec{k},\vec{k},0)$ and
$\Gamma_{2,3}(\vec{k},\vec{k},0)$ normalized to the shot noise
$ N =\langle\hat{S}_0(\vec{k},t)\rangle\sigma$.}
\label{ch:5.fig:9}
\end{figure}

For symmetry reasons Eq.(\ref{eq:gamma1.self}) and
Eq.(\ref{eq:gamma2.self}) give identical results
in the  crossing points of the rings $\vec{k}=\vec{k_H}$
(see Fig. \ref{ch:5.fig:9}).
Since the fluctuations  are isotropically distributed in
$\hat{\bf{S}}$ no "hidden polarization" is observed in these points.
Let us now consider the possibility of quantum effects. First,
the shot noise level for all the Stokes parameters (i.e.
$\Gamma_{i}(\vec{k},\vec{k},0)$ evaluated on coherent states) is
given by the average total intensity
$\langle\hat{S}_0(\vec{k},t)\rangle\sigma$  (see Eq.
(\ref{Eq:It}))\cite{Korolkova.2002,Klyshko-97}.
Inspection of  Eq.(\ref{eq:gamma1.self}) and Eq.(\ref{eq:gamma2.self})
shows that  all the Stokes operators have \emph{classical} statistics
in any FF point, as shown in Fig. \ref{ch:5.fig:9}. In these diagrams
we represent the normal ordered variances. These are
  obtained from the variances Eq.(\ref{eq:gamma1.self}) and
Eq.(\ref{eq:gamma2.self}) after subtraction of the
corresponding shot noise. We see that these are positive quantities
in all the far field,
i.e.  no polarization squeezing appears \cite{Chirkin-93}.
 The
physical reason is that twin photons are emitted with symmetric
wave-vectors (symmetric FF points), while locally no correlations between
orthogonal polarizations are
observed. This motivates us to consider in the next Section
the  correlations between the
Stokes operators of twin beams.

In conclusion, when there is walk-off the polarization state ($P_2$)
varies  in the FF, the fluctuations in the Stokes parameters are above the
classical level in all the far field and no polarization squeezing is
observed.  For $k_y=0$ and in particular for $\pm\vec{k_H}$, $P_2=0$ and
fluctuations are isotropically distributed in all Stokes parameters  so
that the field is completely \emph{unpolarized}. This result would
apply to all the FF plane if there was no walk-off.

\subsection{Far field  correlations}
\label{sect:3.corr}

From Eq. (\ref{eq:gamma1.twin}) and Eq. (\ref{eq:gamma2.twin}) we
see that  $\hat{S}_1$ for symmetric FF modes is {\em
anticorrelated}, while $\hat{S}_2$  and  $\hat{S}_3$ are
positively {\em correlated}. The physical reason for the sign of
these correlations is always the underlying twin photon process
which creates pairs of photons with symmetric wave-vector and
orthogonal polarizations $x$ and $y$, leading to a positive
correlation of the corresponding beam intensities.

These considerations suggest to look for noise suppression in
the following superpositions of Stokes operators
\begin{eqnarray}\label{eq:Sdiff}
D_{1}(\vec{k},\vec{k}',\omega)=\int dt e^{i\omega t}
\{&&\langle[\hat{S}_1(\vec{k},t)+\hat{S}_1(-\vec{k}',t)]\\\nonumber
&&  [\hat{S}_1(\vec{k},0)+\hat{S}_1(-\vec{k}',0)]\rangle\}
\end{eqnarray}
and
\begin{eqnarray}\label{eq:Ssum}
D_{i}(\vec{k},\vec{k}',\omega)=\int dt e^{i\omega t}
\{&&\langle[\hat{S}_i(\vec{k},t)-\hat{S}_i(\vec{k}',t)]\\\nonumber
&&  [\hat{S}_i(\vec{k},0)-\hat{S}_i(\vec{k}',0)]\rangle\}
\end{eqnarray}
with $i=0,2,3$.
We note that $D_{1}(\vec{k},\vec{k}',\omega)=
D_{0}(\vec{k},\vec{k}',\omega)$
(see the definitions Eq. (\ref{eq.S0}) and  (\ref{eq.S1})).
In addition, it follows from
 Eqs.(\ref{eq:gamma2.self})-(\ref{eq:gamma2.twin})  that
 $D_{2}=D_{3}$.
These quantities are non-vanishing for symmetric points
$\vec{k}'=-\vec{k}$.
From the unitarity  of the transformation (\ref{modes_eq}) we know that
$|U_1(\vec{k},\omega)|^2=|U_2(-\vec{k},-\omega)|^2$ and
$|V_1(\vec{k},\omega)|^2=|V_2(-\vec{k},-\omega)|^2$,
so that from Eqs.  (\ref{eq:s2.average}) and (\ref{eq:s3.average}) we
obtain that
\begin{eqnarray}\label{eq:Si.superp.averages}
\langle\hat{S}_1(\vec{k},t)+\hat{S}_1(\vec{k}',t)\rangle
=\langle\hat{S}_i(\vec{k},t)-\hat{S}_i(\vec{k}',t)\rangle=0
\end{eqnarray}
($i=0,2,3$). Therefore Eqs.(\ref{eq:Sdiff}) and (\ref{eq:Ssum})
actually define variances.

From Eq. (\ref{eq:gamma1.self}-\ref{eq:gamma1.twin}) we obtain
\begin{eqnarray}\label{eq:Sdiff.res}
D_{1}(\vec{k},-\vec{k},0)&=&\Gamma_{1}(\vec{k},\vec{k},0)+
\Gamma_{1}(-\vec{k},-\vec{k},0)+\\\nonumber
&&\Gamma_{1}(-\vec{k},\vec{k},0)+
\Gamma_{1}(\vec{k},-\vec{k},0)=0
\end{eqnarray}
for any $\vec{k}$:
There is  complete noise suppression in the sum of the Stokes parameters
$\hat{S}_1$ evaluated in symmetric regions of the transversal field.
Therefore the normal ordered variance is equal to
minus the shot noise  taking nonclassical negative
values.
In conclusion, due to the twin beams intensity correlations, we find
entanglement for $\hat{S}_1$
evaluated in any symmetric FF points and for any pump intensity.

\begin{figure}
\includegraphics[width=0.4\textwidth,clip=true]{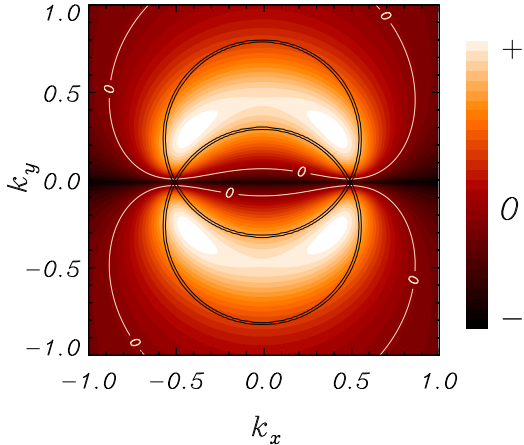}
\caption[$:D_{2}(\vec{k},-\vec{k},0):$]{$D_{2}(\vec{k},-\vec{k},0)$ normal ordered normalized to the
shot noise. The white contour line shows the boundaries between classical
and quantum statistics. The two circles of double continuous line
show the signal and idler maxima intensities.}
\label{ch:5.fig:10}
\end{figure}

\begin{figure}
\includegraphics[width=0.4\textwidth,clip=true]{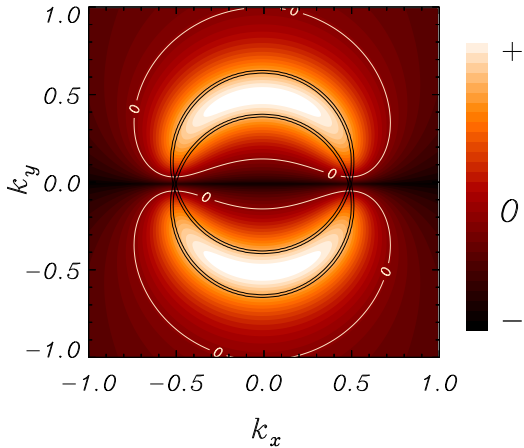}
\caption[$:D_{2}(\vec{k},-\vec{k},0):$ for $\rho_2=0.5$.]{The same as in Fig. \ref{ch:5.fig:10}, but with walk-off
$\rho_2=0.5$.}
\label{ch:5.fig:11}
\end{figure}

The Stokes operators generally
cannot be simultaneously measured with infinite precision (see Eq.
(\ref{eq:heis})). Also the superpositions of Stokes operators involved in
 (\ref{eq:Sdiff}) and
(\ref{eq:Ssum}) are in principle
limited by the Heisenberg relations. However we have that
\begin{eqnarray}
&&\langle[\hat{S}_1(\vec{k},t )+\hat{S}_1(-\vec{k},t ),
 \hat{S}_2(\vec{k},t')-\hat{S}_2(-\vec{k},t')]\rangle=\\\nonumber
&&2i\sigma
 \langle\hat{S}_3(\vec{k},t )-\hat{S}_3(-\vec{k},t )\rangle \delta(t-t')=0
\end{eqnarray}
as follows from Eq. (\ref{eq:Si.superp.averages}).
Therefore there are superpositions of the Stokes operators whose
measurement is not limited by Heisenberg relations because the average of
their commutator vanishes.
In the same way,  also the other superpositions of Stokes operators
 for any couple of symmetric points can be simultaneously measured with total
precision. This result
opens the possibility  to observe noise suppression not only in
the first Stokes operator superposition (\ref{eq:Sdiff.res}), but
also in the other superpositions Eqs. (\ref{eq:Ssum}).
However,  we obtain that
\begin{eqnarray}\label{eq:Ssum.res}
D_{2}(\vec{k},-\vec{k},0)&=&\Gamma_{2}(\vec{k},\vec{k},0)+
\Gamma_{2}(-\vec{k},-\vec{k},0)-\\\nonumber
&&\Gamma_{2}(-\vec{k},\vec{k},0)-
\Gamma_{2}(\vec{k},-\vec{k},0)\neq 0,
\end{eqnarray}
so that $D_{2}(\vec{k},-\vec{k},0)$
generally does not vanish for arbitrary $\vec{k}$.
The profile of the normal ordered $D_{2}(\vec{k},-\vec{k},0)$
is represented in Fig. \ref{ch:5.fig:10}:
we observe that in most part of the far field where there is
large light intensity this quantity is
positive, giving  classical statistics. However
 there is a bandwidth of small
wave-vectors $k_y\sim0$ for which the normal ordered $D_2$ is negative
and therefore quantum effects are observed. This small
bandwidth becomes smaller when increasing the  walk-off, as can be seen
comparing Fig. \ref{ch:5.fig:10} --obtained with $\rho_2=1$-- with
Fig. \ref{ch:5.fig:11} --obtained with $\rho_2=0.5$--.

In particular, along the   ${k}_y=0$ line, we obtain that
$D_{2}({k}_x,-{k}_x,0)=0$. This result  easily follows from
the symmetry  Eq.  (\ref{Eq:symmetry}). Therefore,
along  the ${k}_y=0$ line we have  $D_{1}=D_{2}=D_{3}=0$ which indicates
perfect polarization entanglement between symmetric FF modes. In
summary, in the direction orthogonal to the
walk-off ($k_y=0$), we show  complete noise suppression
in  properly chosen  symmetric modes superpositions of {\bf all} the Stokes parameters.
Along this line the two  points $\pm\vec k_H$ are of
special interest because they have a large photon number.

We finally point out that, the situation studied here is different from
 the case of bright squeezed light
considered in Ref. \cite{Korolkova.2002} and that the perfect
correlations obtained between Stokes parameters  cannot be used
here to obtain an EPR paradox:  in fact from Eq. (\ref{eq:heis})
and Eqs. (\ref{eq:s2.average}-\ref{eq:s3.average}) we obtain that
the Heisenberg principle imposes no limits in the local variances
of the Stokes parameters.

\section{Conclusions}
\label{sect:concl}

We have investigated the  EPR entanglement between
quadrature-polarization components of the signal and idler fields
in symmetric FF points of a type II OPO below threshold, paying
special attention to the effects of walk-off. We have analyzed the
effects of selecting different polarization components: when
walk-off vanishes or in the far field region not affected by
walk-off  ($k_y=0$), there is an almost complete suppression of
noise in the proper quadratures difference of {\em any} orthogonal
polarization components of the critical modes (Sect.
\ref{sect:horizontal}). Selecting non-orthogonal polarization
components the correlations are reduced, vanishing for parallel
polarizations. Walk-off strongly influences the  strength of
correlations. First, the variance of
 the quadratures
difference of orthogonal polarization  components depends  on the reference
polarization: the best EPR entanglement conditions are fulfilled when the
most intense polarization components are locally selected (``vertical
bright" scheme).  If the selected polarizations are not the most intense
(``vertical dark" scheme) we still find EPR entanglement, but correlations
are reduced and there is
 also a rotation of the quadrature angle giving the best
squeezing (Sect.\ref{sect:vertical}).

Our study of EPR quadrature correlations identifies how these
correlations depend on the polarization state.
We have further investigated  non classical polarization properties in terms
of the Stokes operators. The properties of the Stokes parameters in a single
point of the far field do not show any non classical behavior
(Sect.\ref{sect:3}).
For $k_y=0$, where there are no walk-off effects,
 we have shown that the average of the Stokes vector
vanishes, and all the Stokes operators can be measured with perfect
precision. All Stokes operators are very noisy (above the level of coherent
states) and the fluctuations are not sensitive to polarization optical
elements: in fact the field is completely unpolarized, i.e. there is no
 `hidden' polarization.
Quantum effects are observed when considering   polarization
correlations between two symmetric points of the far field.
Still in the direction orthogonal to the  walk-off
($k_y=0$) we show {\em perfect} entanglement of {\em all} the Stokes operators
measured in symmetric FF regions. This result is independent on the
distance to the threshold. These results for $k_y=0$ would apply in all
the FF for vanishing walk-off.
When there is walk-off and for $k_y\neq0$ the entanglement in the second and third Stokes
operators is lost, but for  $\hat{S}_0$ and $\hat{S}_1$ there is still
perfect correlation between two symmetric points of the FF, reflecting the twin photon
process emission.

\section{Acknowledgements}
This work is supported by the European Commission through the
project QUANTIM (IST-2000-26019). Two of us (RZ and MSM) also
acknowledge financial support from the Spanish MCyT project
BFM2000-1108) and helpful discussions with G. Izus and S. Barnett.

\newpage




\begin{thebibliography}{99}
\bibitem{nota}These Stokes parameters are function of the discretized FF modes
introduced in Eq. \ref{Eq:discr_area}.
\bibitem{notabis} The factor
$\sigma\delta_{\vec{k}\vec{k}'}$ appears due to the
discretization introduced for the fields and is
nonvanishing for Stokes parameters measured on
overlapping regions.
\end{thebibliography}
\end{document}